\documentstyle[12pt,aaspp4]{article}
\received{ }
\accepted{ } 
\journalid{337}{15 January 1989} 
\articleid{11}{14}
\def\ds{$\rm D_S$\/}
\def\dc{$\rm D_C$\/}

\def\ltsima{$\; \buildrel < \over \sim \;$}
\def\simlt{\lower.5ex\hbox{\ltsima}} 
\def\gtsima{$\; \buildrel > \over \sim \;$} 

\def\simgt{\lower.5ex\hbox{\gtsima}} 

\def\cm2{cm$^{-2}$\/}
\def\cm3{cm$^{-3}$\/}

\def\o4363{{\sc{[Oiii]}}$\lambda$4363\/}

\begin{document} 
\title{The Close Environment of Seyfert Galaxies and Its Implication for
Unification Models}

\author{D.  Dultzin--Hacyan,  Y.  Krongold, I.  Fuentes-Guridi, \&\ P. Marziani\altaffilmark{1}}
\affil{Instituto de Astronom{\'i}a, UNAM, Apartado
Postal 70--264, M{\'e}xico D.  F.  04510, M{\'e}xico}
\altaffiltext{1}{Osservatorio Astronomico di Padova, Vicolo dell' Osservatorio 5,
I-35122, Padova, Italy}

\begin{abstract}

This paper presents a statistical analysis of the circumgalactic
environment of nearby Seyfert galaxies based on a computer-aided search
of companion galaxies on the Digitized Sky Survey (DSS).  We defined a
sample of 72 nearby Seyfert 1 (redshift 0.007 $\leq$ z $\leq 0.034$) and a
sample of 60 Seyfert 2 galaxies ($0.007 \leq$ z$\leq 0.020$), which include
only high galactic latitude objects.  In addition, we built two control
samples of non-active galaxies matching the number of sample members,
the redshift, morphological type, and diameter distribution of the
Seyfert 1 and Seyfert 2 samples separately. We stress how our sample
selection introduces important methodological improvements that avoid
several sources of strong bias.  An intrinsic difference between the
environment of Seyfert 1 and Seyfert 2 galaxies, suggested by previous
work, is confirmed as statistically significant.  For Seyfert 2
galaxies we find a significant excess of large companions (\dc $\simgt 10
$ Kpc) within a search radius $\simlt$ 100 Kpc of projected linear
distance, as well as within a search radius equal to three times the
diameter \ds\ of each Seyfert galaxy.  For Seyfert 1 galaxies there is
no clear evidence of any excess of companion galaxies neither within
100 Kpc, nor within 3\ds.  For all samples the number of companions
actually counted within a search radius of 3 \ds\ is a factor $\approx$
2 above the expectation values derived from the number density of
galaxies over one square degree fields centered on the sample galaxies,
suggesting a markedly non-Poissonian distribution for galaxies on
scales $\simlt 100$ Kpc. This difference in environment is not
compatible with the simplest formulation of the Unification Model (UM)
for Seyferts: both types 1 and 2 should be intrinsicaly alike, the {\em only}
difference being due to orientation of an obscuring torus.  We propose
an alternative formulation.
\end{abstract}


\section{Introduction}

In the eighties it was found that a relatively
large fraction of Seyferts had a close companion (\cite{dah84,dah85}), although claims that
this excess was due to selection effects were never dismissed
(\cite{fuentes88}).  More recent work revealed significant differences
between Seyfert 1 and Seyfert 2 galaxies (\cite{fins94}), or at least
marginal differences (\cite{der98b}). In both cases an excess of
companions for Seyfert 2 (Sy2) but not for Seyfert 1 (Sy1) with respect
to non-active galaxies.  However, \cite{raf95} also recently, found no
significant difference between Sy1 and Sy2.  We are left in an
uncomfortable situation:  the three most recent and comprehensive works
provide inconsistent results, probably because data are so that the
inherent complexity (and definition ambiguity) of the problem is
starting to affect statistical inferences.  It is not among the aims of
the present paper to toroughly compare all the previous work; this has
been been done by \cite{fins95} and by \cite{dd97}.  The discrepancy
must be however accounted for.  In this work we use for the first time
complete and correctly defined samples, as well as other important
methodological improvements which lead us to confirm the result that it
is only Seyfert type 2 galaxies that have excess companions.  Actually,
there are other indications of intrinsic differences between Sy1 and
Sy2 galaxies.  It has long been known that Sy1 nuclei reside in earlier
morphological type galaxies than Sy2 nuclei.  This has recently been
confirmed from a refined morphological classification of deep HST
images by \cite{malkan98}, along with other morphological differences
which they believe to be intrinsic differences in host galaxy
properties, and which thus undermine one of the postulates for UM.
\cite{dd90} showed that while the mid-infrared ($\sim 25\mu$m) emission
in Sy1 is synchrotron radiation (or dust re-emission of it), in Sy2 it
is dust re-emission of starlight (see also \cite{mouri92}; \cite{dd94};
\cite{gu97}).  In \S \ref{sample} sample selction is described. The
main results are summarized and discussed in \S 3, in \S 4 a brief
discussion of the results is given and finaly, in \S 5 possible
interpretations are analyzed, and some conclusions are drawn.

\section{Sample Selection and Analysis \label{sample}}

The samples of Seyfert galaxies were compiled from the catalog by
 \cite{lipo88}. This catalog was compiled on the basis of the Second
Byurakan Survey (SBS), which is a survey based solely on the UV excess
method. The reason was to avoid the possibility of the inclusion of
Seyfert 2 galaxies serendipitously discovered {\em because} they
belonged to interacting systems (which is the case for some of the
galaxies in the catalog by \cite{veron91}). The importance of
these observational effects were stressed by \cite{marz91}, who also
revealed a fraction of interacting systems larger for Seyfert 2 than
for Seyfert 1.

The present sample consists of 72 Sy1 and 60 Sy2. Both samples are
volume limited, and the $\rm V/V_{{\rm max}}$ test assures uniformity -- and
thus completeness  (Schmidt 1976) -- to a level of 92\%,
The redshifts are limited to $0.007 \leq  z \leq 0.035$ (Sy1) and 
to $0.007 \leq  z \leq 0.020$\ (Sy2), and we
selected galaxies with high galactic latitudes, in order to avoid
extinction, and confusion due to galactic stars. In past work this has
not been properly taken into account. For example, the \cite{fins94}
sample is biased toward low-galactic latitude objects. In \cite{fins94}
the sample contains 53 \%\ of Seyfert galaxies at galactic latitude
$\rm b_{II} \simgt 45^\circ$, while only 27 \%\ are so in the
\cite{raf95} sample. Including low galactic latitude field produces a
bias toward a lower fraction of companion, as detection is more
difficult because of confusion and absorption.  Even if the bias is
equally present in the Seyfert and control samples, a large fraction of
low $\rm b_{II}$\ fields may introduce a bias against intrinsic
differences between Seyfert and control samples.  In this study, we have
selected exclusively galaxies with $\rm b_{II} \simgt 40^\circ$. Also,
rich clusters were avoided.

For the control samples, the above criteria were also imposed. One
important methodological improvement of this work is the definition of
control samples of non-active galaxies which match the Seyfert galaxies
in {\em all} respects except that they are not Seyfert.  In order to
achieve this, {\em two control samples were defined}, one for each type
of Seyfert galaxies, because both the Hubble type and redshift
distributions of the two types of Seyfert galaxies differ.  The control
samples were obtained from a list of more than 10,000 objects of the
CfA catalog (\cite{huchra83}). For each control sample we first matched
Hubble type distributions by artificialy trimming the sample, then we
randomly extracted two subsamples, and proceeded to match redshift
distributions. We did {\em not} match absolute magnitudes since this
would introduce a bias. Matching absolute magnitudes (e.g. De Robertis
et al. 1998) may bias the control sample toward intrinsically higher
luminosity objects, as the Seyfert galaxies host an active nucleus
whose luminosity is expected to be comparable to that of the whole
galaxy. We did match the diameters distribution. Seyfert nuclei reside
most frequently in giant galaxies.  Giant galaxies are relatively rare,
and usually ``do not come alone'': dwarf galaxies are frequently
observed in their immediate surrounding. It is therefore crucial to any
statistical study to have a control sample matching not only redshift
distributions, but the diameters as well as the morphological type of
the Seyfert sample.

The control samples are complete (in volume) to a confidence level of
up to 97\%.  Although the above mentioned similarities were long ago
known to be requiered for a proper comparison (Osterbrock 1993), in
previous works matching the distributions was impossible to achieve
mantaining the same densities, due to the selection of small control
samples from nearby galaxies.  The search for possible  excess of
companions within 100 Kpc is inconsistent with the choice of the
control sample galaxies in the vicinity of the Seyferts (\cite{raf95}; 
\cite{salvato97}).  If Seyferts are at, or close to, the center of
a region of moderate galaxy density enhancement, ``looking around for
the closest non-active spiral galaxy'' means to move (presumably a few
100 Kpc) away from the enhacement and to select areas wich are
systematically of lower density, hence underestimating the fraction of
companions for the control sample, and therefore creating a spurious
excess for Seyfert galaxies.

The procedure to estimate the foreground/background galaxy contamination is as
crucial in this type of statistical work, as is the correct definition of
control samples.  The fraction of Seyfert galaxies with ``physical'' companions
(proximate in space) is the fraction with companions observed within the given
search radius diminished by the fractions of galaxies with an optical companion.
As in previous studies, we derived the probability of finding an optical
companion within a given search radius from the Poisson distributions.

The use of the Lick counts given by \cite{shane} 
to estimate the projection effects can introduce an
important bias (as in \cite{raf95}; \cite{salvato97}; see also
\cite{fins94}). One of the main improvements in this work was the determination
of the number density $\rho$, that goes into the formula for the
predicted number of background galaxies within each area. The
determination was made directly from the DSS plates using FOCAS (Faint Object Classification and
Analysis System; \cite{jarvis81}) to
count galaxies in regions of one square degree surrounding each
galaxy.  In this work the background densities between samples are
statistically equal (according to a Mann-Whitney's U test). Data on
individual objects and on searched sky fields will be presented in a
comprehensive form elsewhere (Krongold et al.  1999, in preparation).

\section{Results \label{results}}

We identified all galaxies with at least one companion within three
times the diameter  of the galaxy (3 \ds). The search was performed
automatically on the DSS with FOCAS, and was limited to galaxies
that could be unambiguously distinguished from stars by the FOCAS
algorithm. This procedure reduces to a minimum several bias present in
previous works, and discussed in the previous section. 

Of 72 Seyfert 1 galaxies $\approx$ 39 \%\ have one companion vs.
$\approx$ 40 \%\ of the 72 galaxies of the Seyfert 1 control sample.
The expected number of optical companions from Poisson statistics is 18
\% and 15 \% for Seyfert 1 and control sample respectively.  If optical
companions are subtracted, then the percentage of galaxies with
presumably physical companions is $\approx 18$\ and $\approx 19$
\%\ for the Seyfert and control sample respectively.  No significant
difference is thus found between the Seyfert 1  and its control
sample.  It is important to stress that the fraction of control sample
galaxies with companions is much higher (a factor $\approx2$) than the
expectation value from Poisson statistics.  Of 60 Seyfert 2 galaxies
$\approx$ 70 \%\ vs 42 \%\ of the control sample show a companion
within 3$\rm D_S$.  The percentage expected from Poisson statistics is
$\approx$ 34 \%\ and $\approx$ 26 \%\ for the Seyfert and comparison
sample.  Thus a large excess (statistically significant to a confidence
level of $\approx$ 99.5 \%) appears to be present for the Seyfert 2
galaxies.  Also in the case of the Seyfert 2 and its control sample,
the fraction of galaxies with companion found is a factor $\approx$2
above the expectation value.  This is an important results of its own
(further discussed in \S \ref{disc}), whichs most likely reflects a strongly
non-Poissonian distributions of galaxies on scales of $\simlt 100$
Kpc.  

\begin{figure} 
\plotone{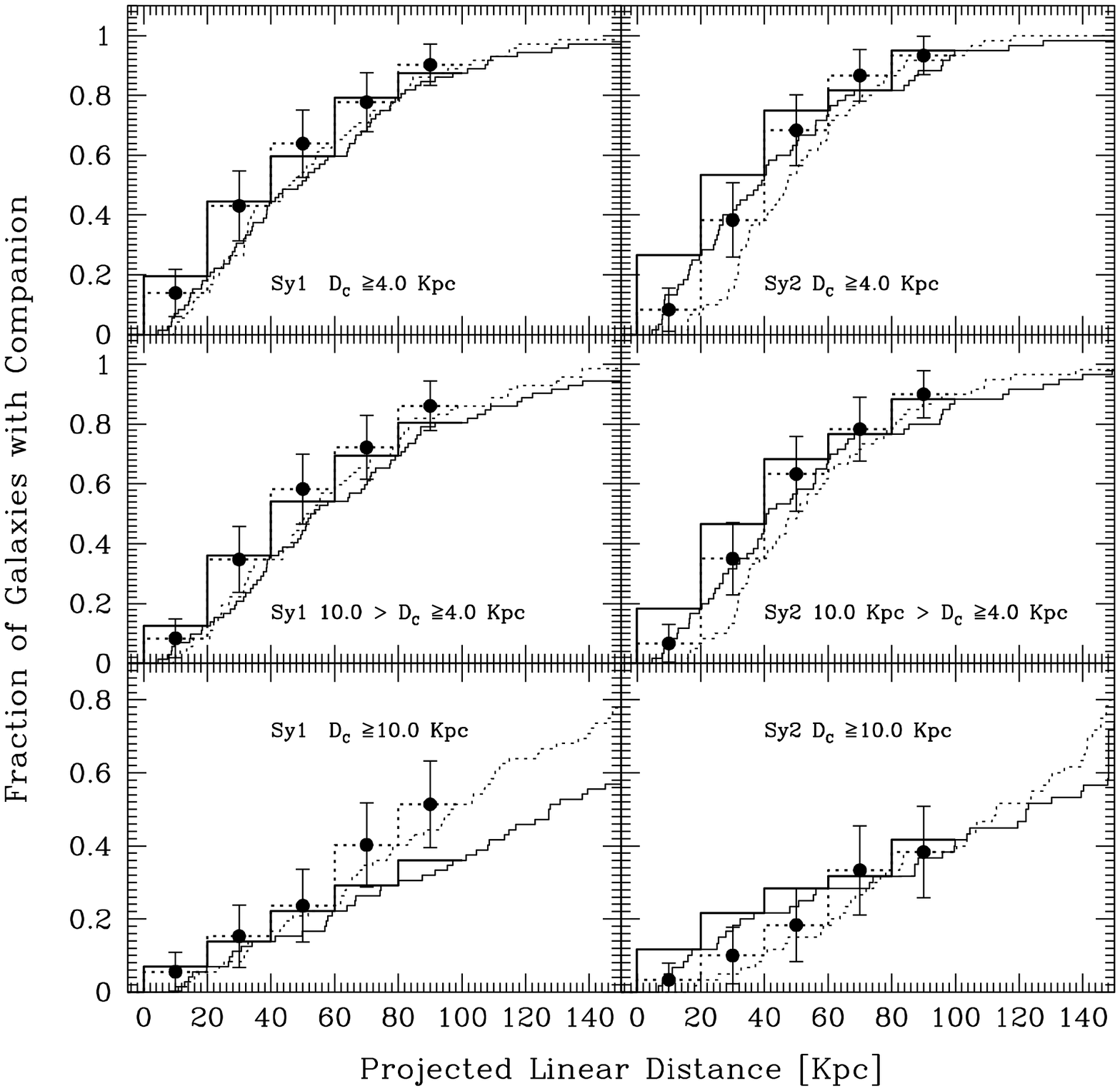}\caption[1]{ Cumulative distribution of projected linear
distances (in units of Kpc) of first companions for Seyfert 1 and control sample
(left panels) and for Seyfert 2 and control sample (right panels).  The uppermost
panels are for all companions (\dc\ $\simgt $ 4 Kpc), the middle ones for
``small'' companions (10 Kpc $\simgt$ \dc $\simgt$ 4 Kpc), and the lowermost ones
are for ``large companions'' (\dc $\simgt$ 10 Kpc).  Dotted lines are for control
samples, filled lines are for Seyferts samples.  The unbinned distribution for
s$\simlt 150 $ Kpc is shown as a thin line, while the thick lines refer to binned
distribution over 20 Kpc bins.  Error bars for the control sample bins are at a 2
$\sigma$\ confidence level.  \label{fig01}} \end{figure}

The cumulative distribution for the projected linear distance \dc\ (in
Kpc) of the first companion is shown in Fig.  \ref{fig01}, without
correction for optical companions.  For these measurements, close
companions were re-identified by eye on the DSS field, and measurements
of centroid position and of diameters were made on computer screen. We
searched for companion galaxies of diameters \dc $\simgt$ 4 Kpc
(assuming $\rm H_0 = 75 km s^{-1} Mpc^{-1}$; this is the limiting
diameter that can be resolved on the DSS by eye and by algorithm at
reshift z$\approx$0.030), within a search radius {\em in all cases}
equal or larger than 100 Kpc of projected linear distance (and in any
case $\simlt$ 250 Kpc).  Above the limiting search radius, we assumed a
``non detection'', and a lower limit to the companion distance was set
equal to the search radius. At 50 Kpc we get 48\%\ galaxies with
companions for Sy1 and 66\%\ for Sy2, frequencies which are close to
those obtained with a variable search radius equal to 3 \ds\ . The
three left panels are for Seyfert 1 and the three panels on the right
for Seyfert 2.  The uppermost panels show the distribution for all
detected galaxies, the middle panel for companion galaxies whose
diameter is 10 Kpc $\simgt$ \dc $\simgt$ 4 Kpc, and the lowermost panel
for, large, bright companions (\dc\ $\simgt$ 10 Kpc).  The thin lines
show the cumulative, unbinned distributions for Seyferts (filled line)
and for control samples (dotted lines).  The distributions binned over
20 Kpc is shown up to a projected linear distance of 100 Kpc (i.e., no
lower limits are included). The error bars on the binned control sample
frequencies were set with a ``bootstrap'' technique (\cite{efron93}) by
randomly resampling the control galaxies into a large number of
pseudo-control samples (3000), and by taking an uncertainty equal to
twice the standard deviation of the distribution of companion frequency
among the pseudo-control samples.

For the binned distribution (thick lines) a marginal statistical
difference is present for small companions within 20 Kpc, and for large
companions if the search radius is extended up to 100 Kpc.  The
situation for Seyfert 2 galaxies is markedly different.  The unbinned
distributions for all galaxies \dc\ $\simgt 4$ Kpc is significantly
different (at a 98 \%\ confidence level), while for small companions
the distributions are only marginally statistically different.  Thus
the difference is driven by a higher frequency of close, large
companions, as shown by the lowermost panel on the right in Fig.
\ref{fig01}.  For companions with \dc\ $\simgt$ 10 Kpc, the difference
in the binned distribution is statistically significant up to $\approx$
60 Kpc.

Summing up, our analysis shows an excess of bright companions within a
search radius of $\approx$ 60 Kpc (or 3 \ds) for Seyfert 2 but not for
Seyfert 1 galaxies, and an excess of galaxies in the close surrounding
of both control and Seyfert galaxies with respect to the expectation of
Poisson statistics.

\section{Discussion \label{disc}}

All studies based on the DSS (including obviously this one) have
limitations intrinsic to the data:  on the DSS, there is a bias against
low surface brightness galaxies at one end, and against compact
galaxies at the other end.  To make things worse, several authors
applied a ``sharp mask'' to the data, ignoring the environment beyond a
search radius three times the diameter of the Seyfert galaxies,
labeling each Seyfert galaxies as ``with'' or ``without'' companions
actually disregarding the complexity and the richness of fields around
several Seyfert galaxies.  A search radius of 3 \ds\ is variable from
object to object:  this may introduce an additional bias that is not
controlled.  In addition, even recent
studies have been based on computer-unaided measurements on the plates
(\cite{fins94}), or worse, on printed enlargements (\cite{raf95}).

The use of the \cite{shane} counts has provided an estimate of the
number of optical companions expected by chance alignment with
background (or foreground) galaxies.  The probability of a chance
alignment within a given search radius is assumed to follow a Poisson
distribution.  However, as clearly shown in this study, the actual
distribution of galaxies within 100 Kpc is markedly non-Poissonian:
control sample galaxies show a much higher fraction (a factor
$\approx$2) of companions than expecteded solely on the basis of
Poisson statistics.  This result is especially robust since counts were
performed over one square degree around the Seyfert galaxies on the
DSS, and is consistent with the observation that, in samples of
perturbed and interacting galaxies (like Vorontsov-Velyaminov's) or
galaxy pairs (like Karashentsev's), the fraction of Seyfert galaxies
appears to be comparable or lower than that expected for fields
galaxies (see the thorough analaysis in \cite{fins95} for
references):  clearly only a minority of interacting systems shows
Seyfert-type activity.  In other words, gravitational interaction may
be a sufficient conditions for activity, but it is certainly not a
necessary condition.  

It is important to stress, that in spite of the above mentioned
limitations, the results obtained by both \cite{salvato97}, and
\cite{der98a} are {\em  not} actualy in contradiction with
the results of the present work. \cite{der98a} conclude
that ``while the companion frequency for Sy2 galaxies is formally
higher, the result is not statisticaly significant (though it is in the
same sense as \cite{fins95}).'' Moreover, they do find that
the mean environment of Sy1 is different from that of Sy2 at a grater
that 95\%\ confidence level, from spatial covariance amplitude
analysis.  The marginal statistical significance of the  differences
found by \cite{der98a}, are in our opinion, just due to small number statistics,
as they are confirmed by  \cite{fins94}, and by the present work which avoids several
sources of bias.

\section{Conclusions}

We confirm an important, disturbing result:  Seyfert 2 galaxies have an excess
of nearby companions while Seyfert 1 do not.  And if Seyfert 1 do not, what
about quasars?  The evidence provided till now about the occurrence of quasars
in interacting host galaxies is based on studies of a few objects and not
statistically significant.  We must stress, however, that this study and the
previous ones address a small subset of interaction phenomenologies that could
give rise to accretion toward a galaxy nucleus:  either we are studying bound
systems or in a stage in which the two galaxies are sufficently close but not
yet merging:  evolved mergers can be well classified as isolated galaxies from
the DSS, or unbound encounters, with separation $\simlt 100$ Kpc.  In both cases we
are considering galaxies whose diameters is $\simgt$ 4-5 Kpc:  it is not
possible to perform a search (with recognition either by eye or by algorithms
implemented on a computer) of smaller galaxies up to z$\approx$ 0.03 without
introducing a redshift-dependent bias.   Morphological disturbances in the inner galactic disks,
which may have been produced in a very close encounter with a small companion,
are  not detected efficiently on the DSS, and were obviously not looked at.  Hyperbolic encounters can have the companion projected
further away than the limiting search radius in a time $\rm t_{fly-by} \sim 0.9 \times
10^8 s_{100 Kpc} v_{1000 km/s}$ yr.
The limitation to three diameters, which corresponds to 60--80 Kpc,
is likely to be inadequate:  the enhancement may be genuinely
restricted to $\simlt$ 100 Kpc, or may be due to a larger density of
galaxies over a larger scale:  on scales $\approx$ 1 Mpc peculiar
motions with respect to the Hubble flow are expected to dominate.  If
we think of the Local Cluster, we realize that a reasonable search
radius should be indeed $\simgt$ 500 Kpc.  Studies with a large search
radius ($\sim$ 500 Kpc) and involving faint companions -- which cannot be carried out on photographic material -- 
have yet to be
done.  

The role of interactions in the induction of nuclear activity is a
complex and open issue. It is particularly  difficult to disentangle  the
differences involved in  ``monster fueling'' and/or circumnuclear
starburst triggering (see e.g. the recent discussion in \cite{der98a}). 
\cite{moles95} investigated all the Seyfert and LINER
galaxies with known morphology, and found that they are all in
interaction or have non-axisymmetric distortions usually with bars,
and/or rings or both. The response to non-axisymmetric perturbations,
is also known to be dependent on the buldge-to-disk ratio. And we must
remember that Sy1 nuclei tend to reside in earlier Hubble type galaxies
thatn Sy2 nuclei, and both types in earlier types than Starburst nuclei
(\cite{ter87}).

Does the difference between the environment of Seyfert 1 and 2 galaxies revealed
in this and in previous studies pose a challenge to unification scheme for
Seyfert galaxies?  In its simplest form, the answer is yes.  A ``minimalist''
interpretation would require to see Sy2 as obscured Sy1 {\em because of
interaction}:  strong interaction with a comparably sized companion enhances
overall star formation, drives molecular gas toward the center of the galaxy,
which may in turn obscure the active nucleus' BLR.  If an ``obscuring torus
scenario'' applies, and if sources are observed at random orientation, then
almost all interacting Sy2 should be obscured Sy1.  This
interpretation allows for an observational verification:  spectropolarimetry of
interacting Sy2 galaxy should reveal a``hidden'' BLR in the majority of
cases.  As only $\sim$ 1/3 of Sy 1 has a companion, this implies that about 2/3
of Sy1 should be genuinely unobscured objects.  An alternative scheme was proposed by
\cite{dd95}:  radiation due to accretion unto a black hole (BH) decreases, while
the relative contribution of a circumnuclear starburst (SB) radiation increases
from Seyfert nuclei types 1 to 2.  Intermediate types can be obviously explained
due to intermediate proportions of these contributions.  Statistical studies of
the multifrequency emission of Seyferts (\cite{mas95}; \cite{dd96}), independently support this scheme.  It is also strongly
supported by direct observations which show that Sy2 galaxies have more
circumnuclear star-forming regions than Sy1, both in the optical
(\cite{gonza97}), and in the IR (\cite{mai95};
\cite{mai97}).  Both alternatives are actualy complemenmtary, since it is the
interaction that drives the needed obscuring and/or star-forming material to the
nucleus.  One thing is clear, these views do not deny the possibility that some,
even the majority of Sy2 galaxies are obscured Sy1.  But an ``only orientation"
difference between Seyfert types is not sustainable.

\acknowledgements

This work was supported by grant IN109896 from DGAPA-UNAM and Italian Ministry for University Research (MURST) under grant Cofin98-02-32

\newpage

\begin{thebibliography}{}

\bibitem[Dahari 1984]{dah84} Dahari, O. 1984, \aj, 89, 966 

\bibitem[Dahari 1985]{dah85} Dahari, O. 1985, \aj, 90, 1772 

\bibitem[De Robertis, Yee, \&\ Hayhoe 1998]{der98b} De 
Robertis, M. M., Yee, H. K. C., \& Hayhoe, K. 1998, \apj, 496, 93 

\bibitem[De Robertis, Hayhoe, \&\ Yee 1998]{der98a} De 
Robertis, M. M., Hayhoe, K., \& Yee, H. K. C. 1998, \apjs, 115, 163 

\bibitem[Dultzin-Hacyan 1995]{dd95} Dultzin-Hacyan, D.  
1995, Revista Mexicana de Astronomia y Astrofisica SC, 3, 31 

\bibitem[Dultzin-Hacyan \& Benitez 1994]{dd94} 
Dultzin-Hacyan, D. \& Benitez, E. 1994, \aap, 291, 720 

\bibitem[Dultzin-Hacyan \& Ruano 1996]{dd96} Dultzin-Hacyan, 
D. \& Ruano, C. 1996, \aap, 305, 719 

\bibitem[Dultzin-Hacyan, Masegosa, \& Moles 1990]{dd90} 
Dultzin-Hacyan, D., Masegosa, J., \& Moles, M. 1990, \aap, 238, 28 

\bibitem[Dultzin-Hacyan et al.  1999]{dd97} Dultzin-Hacyan, D., Krongold, Y.,
Fuentes-Guridi, I., \&\ Marziani, P.  1999, in ``Structure and Kinematics of
Quasar Broad Line Regions'' (San Francisco: ASP CS), in press 

\bibitem[Efron \&\ Tibshirani 1993]{efron93} Efron, B. \&\ Tibshirani, R. 
1993, Introduction to Bootstrap (NY: Chapman \&\ Hall)

\bibitem[Fuentes-Williams \& Stocke 1988]{fuentes88} 
Fuentes-Williams, T. \& Stocke, J. T. 1988, \aj, 96, 1235 

\bibitem[Gonzalez-Delgado \& Perez 1993]{gonza97} 
Gonzalez-Delgado, R. M. \& Perez, E. 1997, 

\bibitem[Gu et al. 1997]{gu97} Gu, Q.S., Huang, J.H., Su, 
H.J., \& Shang, Z.H. 1997, \aap, 319, 92 

\bibitem[Huchra, Davis, \& Latham 1983]{huchra83} Huchra, J., 
Davis, M., \& Latham, D. 1983, Cambridge: Smithsonian Center for 
Astrophysics, 1983,  

\bibitem[Jarvis \& Tyson 1981]{jarvis81} Jarvis, J. F. \& Tyson, 
J. A. 1981, \aj, 86, 476 

\bibitem[Laurikainen \& Salo 1995]{fins95} Laurikainen, E. \& 
Salo, H. 1995, \aap, 293, 683 

\bibitem[Laurikainen et al. 1994]{fins94} Laurikainen, E., 
Salo, H., Teerikorpi, P., \& Petrov, G. 1994, \aaps, 108, 491 

\bibitem[Lipovetsky, Neizvestny, \& Neizvestnaya 1988]{lipo88} 
Lipovetsky, V. A., Neizvestny, S. I., \& Neizvestnaya, O. M. 1988, 
Soobshcheniya Spetsial'noj Astrofizicheskoj Observatorii, 55, 5 

\bibitem[Maiolino \& Rieke 1995]{mai95} Maiolino, R. \& 
Rieke, G. H. 1995, \apj, 454, 95 

\bibitem[Maiolino et al. 1997]{mai97} Maiolino, R., Ruiz, M., 
Rieke, G. H., \& Papadopoulos, P. 1997, \apj, 485, 552 

\bibitem[Malkan, Gorjian, \&\ Tam 1998]{malkan98} Malkan, M. A., 
Gorjian, V., \& Tam, R. 1998, \apjs, 117, 25 

\bibitem[Marziani 1991]{marz91} Marziani P., 1991, Ph D Thesis, SISSA, Trieste

\bibitem[Mas-Hesse et al. 1995]{mas95} Mas-Hesse, J.M., 
Rodriguez-Pascual, P.M., Sanz Fernandez De Cordoba, L., Mirabel, I.F., 
Wamsteker, W., Makino, F., \& Otani, C. 1995, \aap, 298, 22 

\bibitem[Moles, Marquez, \& Perez 1995]{moles95} Moles, M., 
Marquez, I., \& Perez, E. 1995, \apj, 438, 604 

\bibitem[Mouri \& Taniguchi 1992]{mouri92} Mouri, H. \& 
Taniguchi, Y. 1992, \apj, 386, 68 

\bibitem[Rafanelli, Violato, \& Baruffolo 1995]{raf95} 
Rafanelli, P., Violato, M., \& Baruffolo, A. 1995, \aj, 109, 1546 

\bibitem[Salvato \& Rafanelli 1997]{salvato97} Salvato, M. \& 
Rafanelli, P. 1997, Astronomische Nachrichten, 318, 237 

\bibitem[Shane \& Wirtanen 1967]{shane} Shane, C. D., \&\ Wirtanen, 
C. A. 1967, Publ. Lick Obs., Vol. 22, Part I, p. 647 (Chicago:
University of Chicago Press)

\bibitem[Terlevich, Melnick, \& Moles 1987]{ter87} Terlevich, 
R., Melnick, J., \& Moles, M. 1987, IAU Symposia, 121, 499 

\bibitem[Veron-Cetty \& Veron 1991]{veron91} Veron-Cetty, M.-P. 
\& Veron, P. 1991, ESO Scientific Report, Garching: European Southern 
Observatory (ESO), 1991, 5th ed.,  

\end{thebibliography}
\end{document}